%% file: main.tex
\documentclass[sigconf]{acmart}

\input{imports-and-commands}

\AtBeginDocument{%
  \providecommand\BibTeX{{%
    \normalfont B\kern-0.5em{\scshape i\kern-0.25em b}\kern-0.8em\TeX}}}

\copyrightyear{2021} 
\acmYear{2021} 
\setcopyright{acmlicensed}
\acmConference[CHI '21]{CHI Conference on Human Factors in Computing Systems}{May 8--13, 2021}{Yokohama, Japan}
\acmBooktitle{CHI Conference on Human Factors in Computing Systems (CHI '21), May 8--13, 2021, Yokohama, Japan}
\acmPrice{15.00}
\acmDOI{10.1145/3411764.3445400}
\acmISBN{978-1-4503-8096-6/21/05}

\begin{document}

\title{Collecting and Characterizing Natural Language Utterances for Specifying Data Visualizations}

\author{Arjun Srinivasan}
\authornote{This work was conducted while the author was affiliated with the Georgia Institute of Technology.}
\affiliation{%
  \institution{Tableau Research}
  \city{Seattle}
  \state{WA}
  \country{USA}
}
\email{arjunsrinivasan@tableau.com}

\author{Nikhila Nyapathy}
\affiliation{%
  \institution{Georgia Institute of Technology}
  \city{Atlanta}
  \state{GA}
  \country{USA}
}
\email{nikhila.nyapathy@gmail.com}

\author{Bongshin Lee}
\affiliation{%
  \institution{Microsoft Research}
  \city{Redmond}
  \state{WA}
  \country{USA}
}
\email{bongshin@microsoft.com}

\author{Steven M. Drucker}
\affiliation{%
  \institution{Microsoft Research}
  \city{Redmond}
  \state{WA}
  \country{USA}
}
\email{sdrucker@microsoft.com}

\author{John Stasko}
\affiliation{%
  \institution{Georgia Institute of Technology}
  \city{Atlanta}
  \state{GA}
  \country{USA}
}
\email{john.stasko@cc.gatech.edu}

\renewcommand{\shortauthors}{Srinivasan, et al.}

\input{sections/00-abstract}



\maketitle

\input{sections/01-introduction}
\input{sections/02-related-work}
\input{sections/03-study}
\input{sections/04-results}
\input{sections/05-applications}
\input{sections/06-discussion}
\input{sections/07-conclusion}

\input{sections/acknowledgement}

\bibliographystyle{ACM-Reference-Format}
\bibliography{references}





\end{document}

%% file: imports-and-commands.tex
\usepackage{graphicx}

\usepackage{fontawesome}
\usepackage{xcolor}
\usepackage{tcolorbox}
\usepackage{wrapfig}
\usepackage{booktabs}
\usepackage{tabularx}
\usepackage[normalem]{ulem}
\usepackage{enumitem}
\usepackage{amsmath,amssymb}
\usepackage{array}
\usepackage{multirow}
\usepackage{comment}
\usepackage{ifthen}

\definecolor{urlblue}{HTML}{067DE9}

\definecolor{attrRefColor}{HTML}{2C96F3}
\definecolor{chartRefColor}{HTML}{F9A002}
\definecolor{encodingRefColor}{HTML}{9575CD}
\definecolor{aggRefColor}{HTML}{4DB6AC}
\definecolor{designRefColor}{HTML}{F48FB1}

\newtcbox{\buttonTag}{nobeforeafter, colback=white, colframe=gray, boxrule=0.5pt, arc=1pt, boxsep=0pt,left=2pt,right=2pt,top=1.5pt,bottom=1.5pt,tcbox raise base}

\newcommand{\arjun}[1]{\textcolor{black}{#1}}

\newcommand{\added}[1]{\textcolor{black}{#1}}

\newcommand{\projectURL}{\href{https://nlvcorpus.github.io}{\textcolor{urlblue}{\texttt{https://nlvcorpus.github.io}}}}

\newcounter{diffmode}

%% file: sections/00-abstract.tex
\begin{abstract}
Natural language interfaces (NLIs) for data visualization are becoming increasingly popular both in academic research and in commercial software.
Yet, there is a lack of empirical understanding of how people specify visualizations through natural language.
We conducted an online study (\textit{N} = 102), showing participants a series of visualizations and asking them to provide utterances they would pose to generate the displayed charts.
From the responses, we curated a dataset of 893 utterances and characterized the utterances according to (1) their phrasing (e.g., commands, queries, questions) and (2) the information they contained (e.g., chart types, data \added{aggregations}).
To help guide future research and development, we contribute this utterance dataset and discuss its applications toward the creation and benchmarking of NLIs for visualization.
\end{abstract}

%% file: sections/01-introduction.tex
\section{Introduction}

Natural language (NL) is gaining traction as an input modality for data visualization tools.
Both mainstream visualization systems (e.g., Microsoft Power BI~\cite{mspowerbi}, Tableau~\cite{tableauaskdata}) and research prototypes (e.g.,~\cite{sun2010articulate,gao2015datatone,setlur2016eviza,yu2019flowsense,cui2019text,srinivasan2020interweaving}) have demonstrated the potential natural language interfaces (NLIs) hold for supporting a rich visual analytic flow and to cater to broader audiences (e.g., visualization novices, people with disabilities).
\textit{Visualization specification} plays a central role in these tools, as people create charts to visually explore and analyze their data (e.g., creating histograms to check data distributions, creating scatterplots to observe correlations), as well as to communicate their findings with other stakeholders (e.g., creating grouped bar charts for quarterly reports).

However, beyond the space of utterances supported in current NLIs for visualization, there is limited empirical understanding about the nature of utterances people use to specify data visualizations through NL.
For instance, how do people structure or phrase their utterances?
Do they use systemic commands (e.g., \textit{``Plot scatterplot of sales by profit.''}) or high-level questions (e.g., \textit{``Have cars gotten lighter over time?''})?
What type of information do people include in their utterances?
Do they explicitly list chart types and aggregations (e.g., \textit{``Bar chart showing average profit by state''}) or expect systems to infer such information for them (e.g., \textit{``Visualize profit across states.''})?
Furthermore, in cases where people provide explicit references, are these references direct (e.g., exact or partial matches to attribute names), or implied or semantic (e.g., using synonyms for attribute names, using phrases like \textit{``How many''} instead of specifying a \texttt{COUNT} aggregation)?
Answering these questions can help assess how well-aligned current tools are with people's expectations from NLIs for visualization.
This, in turn, can help developers of NLIs for visualization validate and boost their systems' performance and improve system usability.

We conducted an online study with 102 participants, curating a dataset of 89\added{3} visualization specification-oriented NL utterances.
Specifically, we showed participants a series of ten canonical visualizations (e.g., bar charts, line charts, scatterplots) and asked them to enter utterances they would use to create the shown charts.
We then characterized the resulting utterances along two dimensions: 1) their phrasings, i.e., how the utterances were structured (e.g., questions, commands), and 2) the information contained in them (e.g., some utterances explicitly requested a visualization type).
For the remainder of this paper, we use the term \textbf{utterance} to refer to any NL command, statement, query, question, or instruction that one may issue to an NLI.

Leveraging this characterization, we provide a high-level summary of the types of utterances that people use to specify visualizations, discussing their coverage within current systems and revealing underexplored classes of utterances.
\added{We also distill our experiences and highlight critical considerations for designing and conducting online studies to collect data for NLIs for visualization.
Finally, we discuss how the curated dataset of utterances can be used toward the creation and benchmarking of NLIs for visualization, and for collecting larger datasets.}

The key contributions of this work are twofold: (1) a publicly available dataset of 89\added{3} visualization specification-oriented utterances and a discussion of its application toward creating and benchmarking NLIs, and (2) \added{a characterization of NL utterances people use to specify data visualizations along with its implications for designing future NLIs for visualization specification}.


%% file: sections/02-related-work.tex
\section{Related Work}



Advances in NL recognition and understanding have led to a surge of interest in NLIs for data visualization.
In recent years, a wide range of systems have been developed to support NL interaction during visual data exploration (e.g.,~\cite{sun2010articulate,gao2015datatone,setlur2016eviza,hoque2017applying,srinivasan2020inchorus,srinivasan2020interweaving}), for question answering with charts~\cite{kim2020answering}, and for facilitating data-driven communication~\cite{cui2019text,lai2020automatic}.
A majority of these systems focus on specific visualizations and usage scenarios (e.g., network data exploration~\cite{srinivasan2017orko}, dataflow diagram editing~\cite{yu2019flowsense}) or technical aspects of system development (e.g., detecting ambiguities~\cite{gao2015datatone,setlur2016eviza}, inferring underspecified queries~\cite{setlur2019inferencing}).

On the other hand, some projects have more closely investigated the structure of NL utterances and thus are highly relevant to our work. 
Metoyer et al.~\cite{metoyer2012understanding} conducted a study where 20 participants worked in pairs with one participant verbally describing a given chart to another participant who attempted to draw the chart based on that description.
Through the collected data, the authors discuss implications for future NL systems including the need to support ambiguous instructions (e.g., \textit{``a narrow bar''}) and the use of relative terms to arrange spatial layouts (e.g., \textit{``the yellow bar is stacked on top of the blue bar''}), among others.
Setlur et al.~\cite{setlur2016eviza} conducted a study to collect utterances people may pose when interacting with a given chart.
They identified different task categories people try to perform (e.g., search, filter, change chart type) and used the corresponding utterances to implement the Eviza system~\cite{setlur2016eviza}.
Tory and Setlur~\cite{tory2019mean} conducted a wizard-of-oz study to understand people's expectations of the role of system intelligence in NLIs during visual analysis.
Their findings highlight that NLIs can enable a richer visual analytic flow by supporting both explicit and implicit user intent, by prioritizing explicit intent over consistency in use of graphical encodings during chart transitions, and by performing proactive actions, such as visually encoding filters and implicitly applying data transformations, among others.
Our work shares a similar goal with these studies in that we also seek to understand the nature of NL utterances within the context of data visualization.
However, we focus on collecting and understanding utterances people may pose to a system for visualization specification (as opposed to interacting with a given chart).
In doing so, we provide additional evidence for findings from prior studies while also offering novel insights into classes of utterances not supported in current tools and discussing practical implications for system design.

Recent work has also begun to curate datasets of utterances to enable the use of machine learning techniques for NLI development.
For instance, during an online study, Kim et al.~\cite{kim2020answering} showed people a series of bar charts and line charts, and collected questions people pose to these charts along with the participants' answers (and explanations) to those questions.
This data was then used to update Sempre~\cite{berant2013semantic}, a model conventionally used for question answering within relational data tables, to answer questions in the context of data visualizations.
Fu et al.~\cite{fu2020quda} collected utterances mapping to Amar et al.'s 10 low-level analytic tasks (e.g., filter, sort, find extremum)~\cite{amar2005low}.
Using the collected data to train a text classification model, the authors showed how one could detect analytic tasks from given NL utterances.
Similar to these recent studies, we also aim to foster future research and development by curating a freely available dataset of utterances.
To this end, we complement the existing datasets that focus on question answering and task detection, and contribute the first dataset that contains visualizations and utterances that were used to specify those visualizations.
Besides discussing the applications of the presented dataset, we also offer insights into unique challenges we faced during data collection and curation as these can help extend the described study and generate larger datasets.


%% file: sections/03-study.tex
\section{Online Study}
\label{sec:study}

\begin{figure*}[t!]
    \centering
    \includegraphics[width=.9\textwidth]{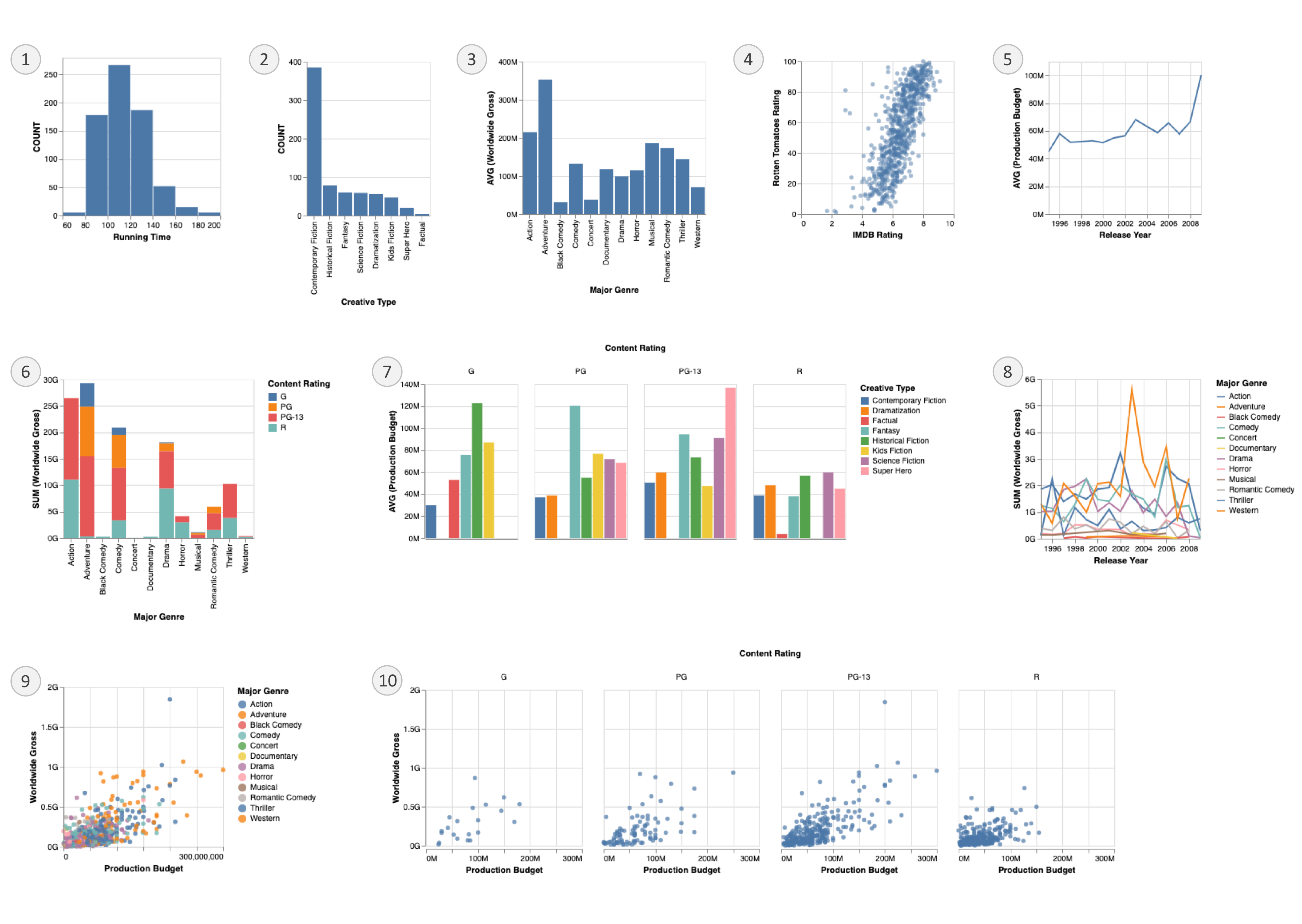}
    \vspace{-1em}
    \caption{Ten visualizations used for the study (all charts were created using Vega-Lite~\cite{satyanarayan2016vega}).
    Charts shown in the figure represent the movies dataset.
    Charts (1, 2) visualize one quantitative or categorical attribute using the \texttt{x}- and \texttt{y}-encoding channels. 
    Charts (3-5) also only encode data via the \texttt{x}- and \texttt{y}-channels but visualize two attributes at a time.
    Finally, charts (6-10) use \texttt{color} and/or \texttt{column} (for faceting) as an additional encoding to visualize three attributes at a time.
    The corresponding ten charts for the cars and superstore data are included in the supplemental materials.
    }
    \label{fig:visualizations-used}
    \Description{The ten visualizations used for the study.}
\end{figure*}


\subsection{Visualizations and Datasets}

\input{tables/dataset-summary.tex} Our study covered a total of 30 visualizations (10 visualizations x three datasets): Figure~\ref{fig:visualizations-used} shows the 10 visualizations that were presented during each session and Table~\ref{tab:dataset-summary} summarizes the three datasets used in the study.
These visualizations covered three popular chart types: \textit{bar charts}, \textit{line charts}, and \textit{scatterplots}, along with their variants (histograms, stacked \& grouped bar charts, multi-series line charts, and colored \& faceted scatterplots).
Our choice of visualizations was motivated by the popularity of these charts across visualization tools \added{in general~\cite{battle2018beagle} as well as within existing NLIs for visualization (e.g.,~\cite{sun2010articulate,gao2015datatone,setlur2016eviza,srinivasan2020inchorus})}.
Furthermore, these visualizations also allowed us to cover three basic data types: \textit{categorical} (both nominal and ordinal values), \textit{quantitative}, and \textit{temporal}, as well as their combinations involving one to three data attributes at a time.
Collectively, this set of visualizations and attribute type combinations ensured that the utterances collected via the study would be broadly applicable and could be used in the context of general visualization tools.

For the data underlying these charts, we decided to use datasets that: (1) contain a mix of categorical, quantitative, and temporal values, and (2) represent a generally understood domain to engage a broad participant pool. 
With these criteria in mind, we chose three tabular datasets covering different domains: \textit{cars}, \textit{movies}, and \textit{superstore sales} (Table~\ref{tab:dataset-summary}).

\subsection{Participants}

To collect utterances that were representative of what end-users of a visualization system might provide, we sought to recruit participants who are actual users of visualization tools or are interested in visual data analysis.
With this target population in mind, we shared the call for participation for the study via mailing lists at universities and software companies.
We specifically used mailing lists that included data systems-related users (e.g., database and visualization courses, dashboard tool user groups, visualization tool developers and product managers).
We also posted the call for participation on public social platforms such as LinkedIn and Reddit.
On these social platforms, we again posted only in targeted interest groups pertaining to data visualization and visual analytics (e.g., `r/visualization' and `r/PowerBI' communities in Reddit, Tableau user groups on LinkedIn, the Data Visualization Society Slack channel).
Participation was voluntary and no compensation was provided (neither financially nor via other reward mechanisms such as course credit).
Participants could choose to exit the study at any point.
Participants were also informed beforehand that the utterances collected through the study will be made publicly available and that no personal information (contact details or demographic information) will be captured during the study.

Over a span of $\sim$60 days, a total of 202 participants visited the study URL among which 102 participants took part in the study (i.e., completed at least one trial).
Seventy-six participants completed the entire session, entering utterances for all ten visualizations while 26 participants exited the study amidst a session.

\subsection{Procedure and Task}

Each session included four major phases: (1) \textit{introduction \& consent}, (2) \textit{dataset \& task description}, (3) \textit{task trials}, and (4) \textit{providing prior experience with visualization tools}, and consisted of 10 trials corresponding to the 10 visualizations displayed in Figure~\ref{fig:visualizations-used}.
Each participant saw all 10 visualizations from only one of the three datasets (cars, movies, superstore).
Both the dataset selected and the order of visualizations were randomized across participants.

\begin{figure*}[t]
    \centering
    \includegraphics[width=\linewidth,keepaspectratio]{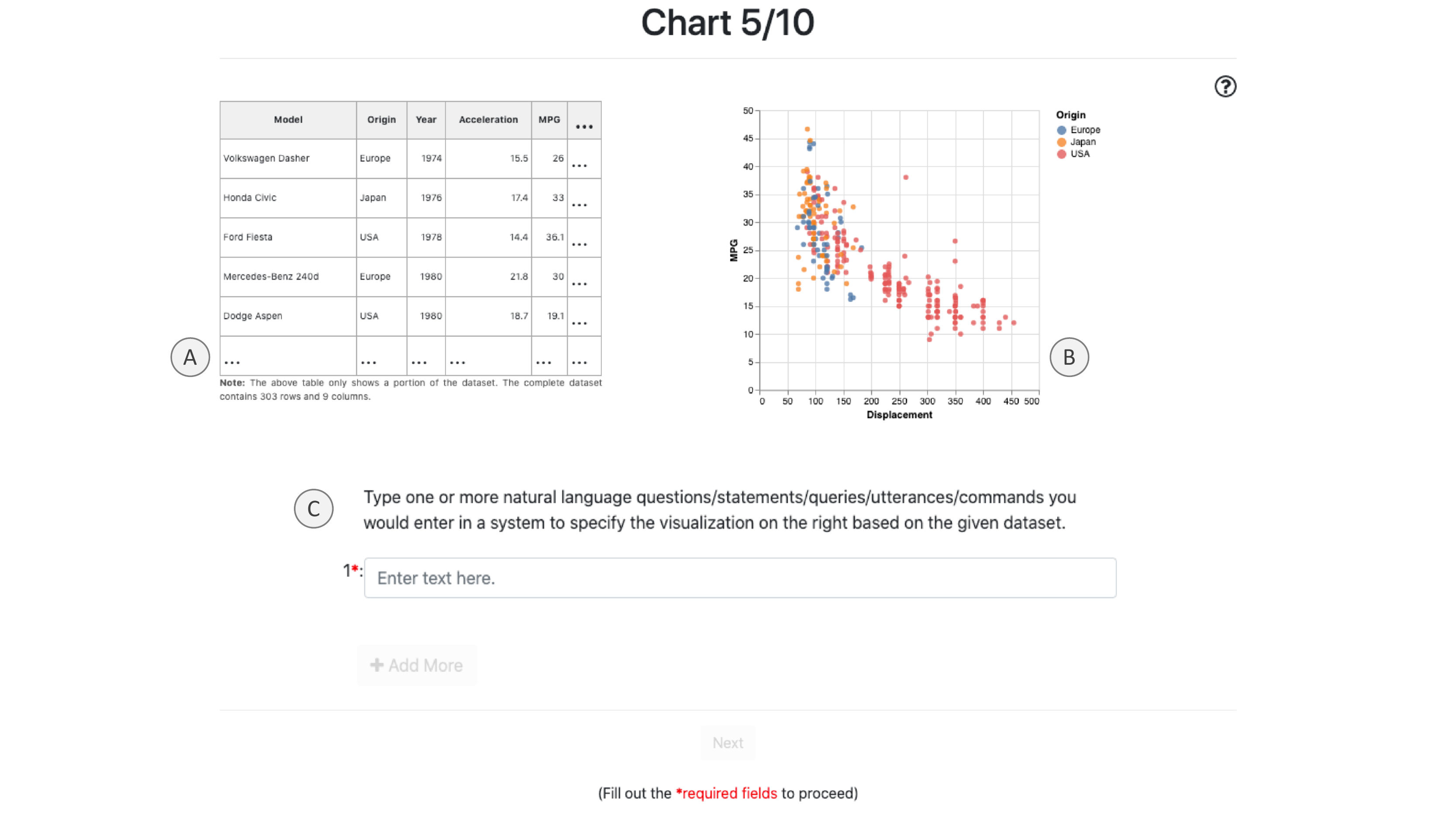}
    \caption{Screenshot from the study illustrating a single task trial. The example shown is from a session visualizing the cars dataset. \added{(A) Dataset preview, (B) chart to specify, and (C) task prompt and input.} Besides entering utterances, participants could also click the {\small{\faQuestionCircleO}} icon on the top-right corner to revisit the dataset and task overview.}
    \label{fig:trial}
    \Description{A screenshot from the study illustrating a single task trial.}
\end{figure*} 

Participants were first informed about the study goal, compensation, public sharing of collected utterances, etc.~and were asked to provide their consent.
Next, participants were given a brief introduction to the dataset (e.g., number of rows and columns, what the rows and columns represent) and the task, which required participants to enter (type in) one or more utterances they would pose to a system like Tableau or Microsoft Power BI to specify a displayed visualization (Figure~\ref{fig:trial}).
Note that to ensure participants were not biased to phrase utterances in a specific way, we did not provide any example utterances as part of the task description.
Participants were then presented with the 10 trials in random order.
Lastly, participants who completed the trials were given the option to provide their prior experience level using visualization tools on a five-point scale ranging from `Never' to `Very Frequently'.
All utterances entered during the study were logged for further analysis.

\subsection{\added{Study Design Considerations}}

\added{We iteratively developed the final study interface and procedure described above through 41 pilot sessions, including two in-person group sessions\footnote{\added{During the group sessions, participants completed the study individually on their laptops and then collaboratively provided feedback on issues they faced (e.g., incorrect chart sizes) or instructions that were confusing to them (e.g., task phrasing).}} (with 11 and 14 participants, respectively) and 16 distributed online sessions.
To the best of our knowledge, this is the first online study designed specifically for collecting NL utterances for visualization specification.
Correspondingly, below we briefly reflect on the pilot feedback and the evolution of our study design to highlight key considerations for future online studies investigating NL input for data visualization.}

\subsubsection{\added{No training trials or sample inputs.}}
\added{As we required participants to enter utterances that naturally occurred to them, we did not present a ``training phase'' or sample inputs at the start of the study to avoid biasing the utterances participants entered.
This absence of training trials is one crucial consideration compared to more traditional perception- and interaction-focused online visualization experiments that use training trials to acclimatize participants to the study tasks and interface.}

\subsubsection{\added{Phrasing task prompts.}}
\added{One important factor for future studies to consider is the phrasing of the task prompt (Figure~\ref{fig:trial}C).
To cite an anecdote, during the pilots, we initially used the prompt ``\textit{Enter one or more natural language commands or utterances that you would pose to a system to create the visualization shown on the right.}''
With this prompt, a majority of the input utterances were phrased along the lines of \textit{``Create a [chart type] of [attributes]''} where the start \textit{``Create a''} was consistent across almost all participants.
While these were natural utterances for some participants, other participants said that the prompt nudged them towards phrasing their utterances in a particular way.
Correspondingly, to repeatedly remind participants that they were free to phrase utterances per their individual preferences, we updated the task prompt to (\textit{``Type one or more natural language...''} as shown in Figure~\ref{fig:trial}C.
Specifically, we included a variety of terms to refer to participant input (e.g., questions/statements/queries/...) and randomized the ordering of these terms (e.g., utterances/queries/statements/..., queries/questions/commands/...), and also varied the verbs used in the prompt (e.g., create, generate, specify).}

\subsubsection{\added{Choosing the number of required utterances.}}
\added{Setting the minimum number of utterances that participants need to enter is also an important study design factor to consider.
We initially asked the pilot participants to enter \textit{two or more} utterances they would use to specify the given chart.
However, participants commented that it was challenging to provide more than one ``natural'' utterance, leading them to create a contrived second utterance by paraphrasing their first utterance.
Furthermore, instead of reading ``\textit{two or more}'' as two mutually exclusive utterances to specify the shown chart, some participants misinterpreted the instructions as a mandate for entering at least two utterances that collectively specified the shown chart (e.g., \textit{``Show me a bar chart''~$>$~``Add states and average profit"}).
We therefore asked participants to enter \textit{one or more} utterances in the final study (Figure~\ref{fig:trial}C).}

\subsubsection{\added{Providing a dataset preview.}}
\added{
One subtle but important design consideration is the inclusion of a dataset preview as part of each trial (Figure~\ref{fig:trial}A).
In the initial pilots, we included the dataset preview only during the introductory phase but not for the actual trials. 
However, we observed that including a dataset preview alongside the chart gave participants a better understanding of the dataset, helping them formulate more naturalistic utterances involving semantic and value-based references to data attributes (e.g., using phrases like \textit{``over time''} or \textit{``movie length''} instead of the attribute names \textit{Year} or \textit{Running Time}, or values like \textit{``furniture, office supplies, and technology''} to refer to \textit{Category}).}

%% file: tables/dataset-summary.tex
\begin{table}[b!]
\centering
\resizebox{\linewidth}{!}{%
\begin{tabular}{@{}lccc@{}}
\toprule
\textbf{\begin{tabular}[c]{@{}l@{}}\Large{Dataset}\\ \Large{(Cardinality)}\end{tabular}} &
  \textbf{\begin{tabular}[c]{@{}c@{}}\Large{Categorical}\\ \Large{Attributes}\end{tabular}} &
  \textbf{\begin{tabular}[c]{@{}c@{}}\Large{Quantitative}\\ \Large{Attributes}\end{tabular}} &
  \textbf{\begin{tabular}[c]{@{}c@{}}\Large{Temporal}\\ \Large{Attributes}\end{tabular}} \\ \midrule
\Large{Cars (303)}        & \Large{2}  & \Large{5} & \Large{1} \\
\Large{Movies (709)}      & \Large{3}  & \Large{5} & \Large{1} \\
\Large{Superstore (5899)} & \Large{11} & \Large{5} & \Large{1} \\ \bottomrule
\end{tabular}%
}
\vspace{.5em}
\caption{Summary of three datasets used for the study (complete datasets are provided as supplementary material).}
\vspace{-2em}
\label{tab:dataset-summary}
\Description{A list of datasets used for the study along with metadata including the number of rows and column types in each dataset.}
\end{table}

%% file: sections/04-results.tex
\section{Analysis and Results}

\subsection{Data Cleaning}
\label{sec:data-cleaning}

A total of 803 logs were generated from the 102 sessions (76 complete + 26 partial sessions, one log per trial).
We manually inspected these logs to curate a set of logs containing only valid utterances that could reliably be used for future research and development.
We discarded 159 trials, leaving a total of 644 trials:
we discarded trials if they contained utterances that fell into the following categories:

\begin{itemize}[leftmargin=.175in]
    \item \textbf{Arbitrary text} (45): Trials that contained utterances composed of an arbitrary collection of characters. Examples include \textit{``njknjk,''} \textit{``xyz,''} \textit{``fdsfs,''} and \textit{``na.''}
    
    \item \textbf{Code or SQL} (43): Trials that contained SQL commands or code in programming languages, such as R and Python. Examples include \textit{``SELECT Year, Origin, AVG(Horsepower) FROM dataset WHERE Origin IN (`Europe', `Japan', `USA') GROUP BY Origin, Year''} and \textit{``ggplot(dataset) + geom\_point(mapping= aes(x= Horsepower, y= Acceleration)''}.
    
    \item \textbf{Incomplete utterances} (37): Trials with utterances that were incomplete in the context of recreating the displayed visualizations.
    For instance, for the trial shown in Figure~\ref{fig:trial}, we excluded the utterance \textit{``show me displacement vs miles per gallon''} because it does not specify, in any way, that \textit{Origin} should be visualized as a third data attribute \added{and used for coloring the points}.
    \added{While incomplete utterances may be considered valid for the given datasets during general data exploration, they lacked sufficient information for a visualization system to adequately infer the shown chart from the utterance alone}.
    Other discarded examples under this category include \textit{``Bar graph''} (for a bar chart that needs two attributes \added{similar to Figure~\ref{fig:visualizations-used}-3}), \textit{``average of horsepower''} (for a multi-series line chart \added{similar to Figure~\ref{fig:visualizations-used}-8}), and \textit{``Show me bar graphs of each ship mode''} (for a stacked bar chart with three attributes \added{similar to Figure~\ref{fig:visualizations-used}-6}).
    
    \item \textbf{Miscellaneous utterances} (34): Trials containing utterances where it looked like participants did not follow the instructions and entered utterances that were not relevant to the task of specifying a displayed chart.
    For example, for the chart in Figure~\ref{fig:trial}, we discarded a trial composed of only the following two utterances: \textit{``Overlay Legend top right''} and \textit{``Add Exponential Trend.}''
    Other discarded examples include \textit{``add data labels, inside end,''} \textit{``Change y-axis major units to 2,''} and \textit{``Title start with A.''}
\end{itemize}



\subsection{Results}

Among the filtered group of 644 trials, 252 corresponded to the cars dataset, 205 to the movies dataset, and 187 to the superstore dataset.
\added{The number of trials for each [dataset x chart type] combination ranged 15--28 (\textit{M} = 21).}
The 76 participants who completed all 10 trials also provided their prior working experience with visualization tools (Never: 3, Somewhat infrequently: 9, Occasionally: 15, Somewhat frequently: 23, Very frequently: 26).
In terms of utterances, from the 644 trials, we extracted a total of 89\added{3} utterances with individual trials containing one to seven utterances.

\vspace{.3em}
\noindent\textbf{Characterizing Utterances.} We qualitatively analyzed the 89\added{3} utterances through an open coding process.
Specifically, two of the authors collaboratively inspected the data to identify high-level themes and generated a set of codes to characterize and group utterances.
We then individually tagged random subsets of utterances using the initial coding scheme and compared the coding results for agreement.
The resulting codes were then collectively discussed and refined for consistency until an 85\% agreement was reached using the Jaccard Index.
Subsequently, all 89\added{3} utterances were tagged using the mutually agreed-upon codes.

\begin{figure}
    \includegraphics[width=\linewidth,keepaspectratio]{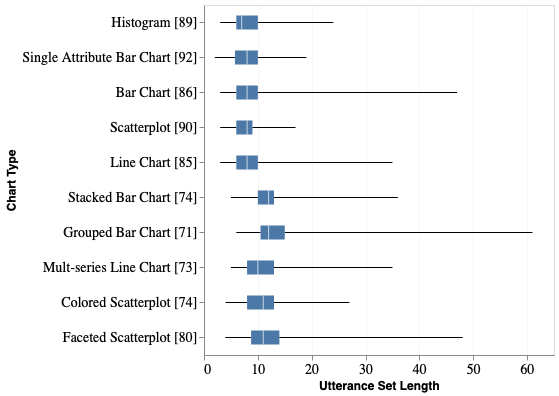}
    \caption{\added{Summary of the lengths of the 614 utterance sets across the 10 chart types (Figure~\ref{fig:visualizations-used}) and three datasets (Table~\ref{tab:dataset-summary}) used in the study.
    The [count] alongside the chart types indicates the number of utterance sets.
    }}
    \label{fig:utterance-set-lengths}
    \Description{A box plot summarizing the lengths of utterance sets across the ten chart types.}
    \vspace{-1em}
\end{figure}

From the 89\added{3} utterances, we identified a total of 814 \textit{\textbf{utterance sets}}: we define an utterance set as a set of one or more utterances that collectively map to a specific visualization.
Utterance sets can either be \textit{singleton} (755) or \textit{sequential} (59).
A singleton utterance set contains only one utterance intended to specify one chart.
Examples of singleton utterance sets for the chart shown in Figure~\ref{fig:trial} include (1) \textit{``Scatterplot mpg vs displacement color by origin''} and (2) \textit{``What is the correlation between displacement and MPG of cars with different origins?''}
On the other hand, a sequential utterance set is composed of multiple utterances that collectively create one chart.
Going back to the chart in Figure~\ref{fig:trial}, an example of a sequential utterance set (with two utterances) is \textit{``plot displacement by mpg''~$>$~``color by origin''} (the symbol~`$>$' denotes a follow-up utterance).
\added{The 59 sequential utterances were composed of 2--5 utterances (\textit{M} = 2).}
\added{The lengths of utterance sets ranged 2--61 words (\textit{M} = 10), with individual utterances containing up to 38 words (\textit{M} = 9), see Figure~\ref{fig:utterance-set-lengths}.}

\begin{figure*}[t!]
    \centering
    \includegraphics[width=.9\textwidth, keepaspectratio]{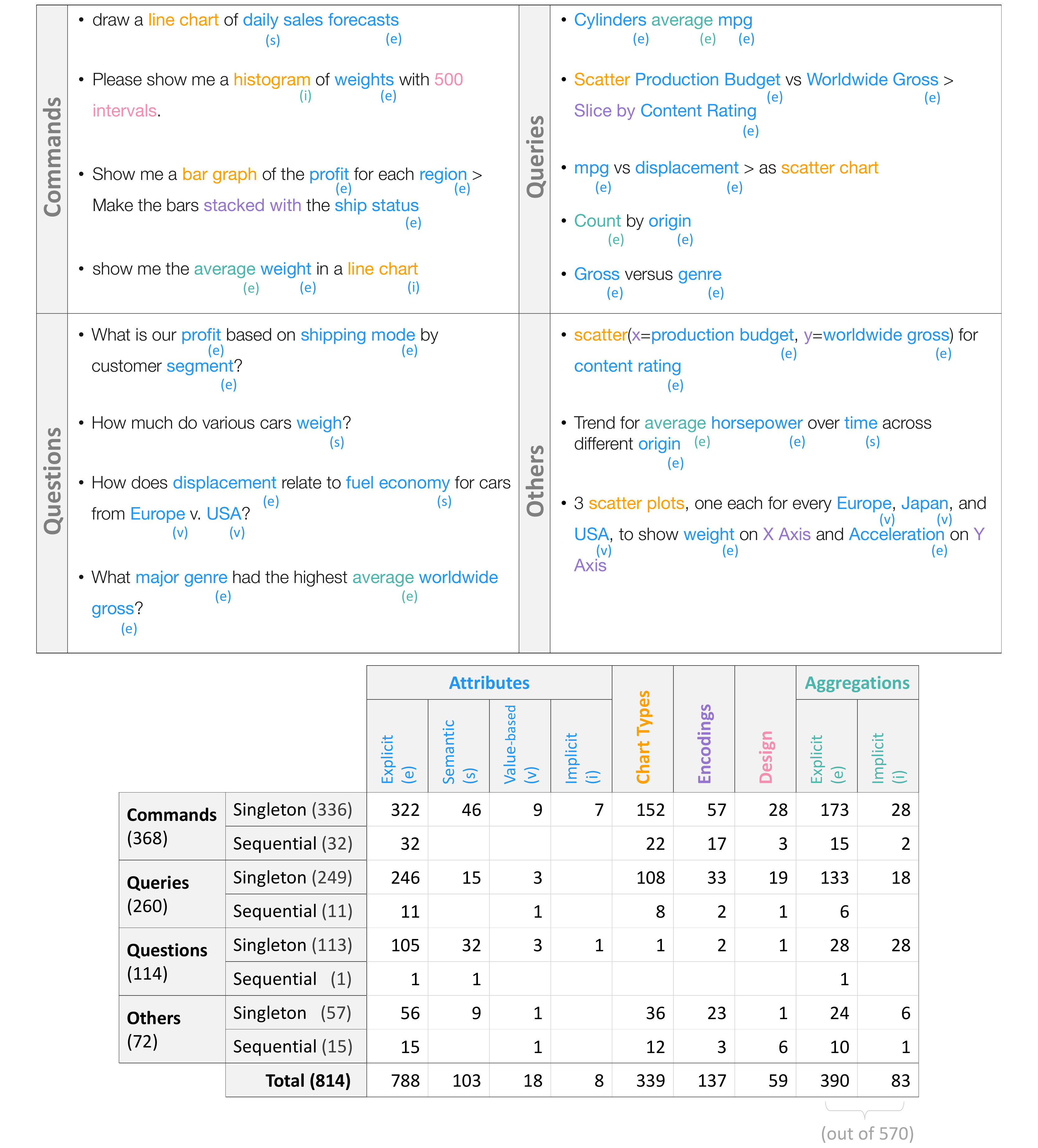}
    \vspace{-1em}
    \caption{\added{An overview of the curated dataset of 814 utterance sets composed of 89\added{3} utterances from the 644 trials. (Top) Utterance phrasing types we observed along with examples of each type. (Bottom) A distribution of information references within each utterance set. The color codes of the columns headers are also used to highlight the corresponding references in examples shown at the top. Note that specifically for attribute references, a single utterance set can have multiple types of references and hence the sub-category totals add up to more than 814 (e.g., \textit{``How does displacement relate to fuel economy for cars from Europe v. USA''} contains an explicit reference to \texttt{Displacement}, a semantic reference to \texttt{MPG}, and a value-based reference to \texttt{Origin}).
    Additionally, data aggregations were only applicable to 570/814 utterance sets that were issued in the context of visualizations other than scatterplots (since scatterplots did not show aggregated values) and 97 of these 570 utterance sets did not include explicit or implicit aggregation references (e.g., ``\textit{bar chart for mpg v/s cylinders}'').}}
    \label{fig:results}
    \Description{An overview of the curated utterances in terms of their phrasings and the information they contained.}
\end{figure*}

To better understand the nature of utterances that people use to specify visualizations, we characterized the 814 utterance sets along two key dimensions: \textsc{\textbf{How}} the utterances are phrased and \textsc{\textbf{What}} dataset or visualization-specific information they contain.
Figure~\ref{fig:results} summarizes the results of this characterization.

In terms of phrasing variations, we tagged each of the 814 utterance sets as one of: \textit{commands} (368), \textit{queries} (260), \textit{questions} (114), or \textit{others} (72).
Figure~\ref{fig:results}-top presents examples of utterances under each of these phrasing categories.
Specifically, we defined `commands' as utterances that were phrased similar to instructions or requests from one person to another.
We tagged an utterance set as a `query' when the underlying utterances were
phrased as keywords or terse web search-like queries.
`Questions' were data-driven questions that participants expected to see the displayed visualization as a response to.
If an utterance set did not fall under one of the three aforementioned phrasing categories, we tagged it as `others' because such phrasings were relatively infrequent.
Broadly speaking, the types of phrasings in the `others' category included caption-like statements \added{describing a chart}, \added{commands that were primarily in NL but also used special characters like `\textit{=}' and `\textit{( )}' that have a programming connotation}, or detailed chart rendering instructions.

With respect to the information contained within utterances, we identified five types of references that would be relevant to an NLI when interpreting utterances to create visualizations: \textit{\textbf{\textcolor{attrRefColor}{attribute}}}, \textit{\textbf{\textcolor{chartRefColor}{chart type}}}, \textit{\textbf{\textcolor{encodingRefColor}{encoding}}}, \textit{\textbf{\textcolor{aggRefColor}{aggregation}}}, and \textit{\textbf{\textcolor{designRefColor}{design}}} references.
Figure~\ref{fig:results}-top highlights these references in the context of sample utterances illustrating the different phrasing variations.
Attribute references are essentially words in a query that map to a data attribute.
Attribute references can be further divided into four sub-categories: (i) explicit references, where words in a query directly match a portion of a data attribute (e.g., \textit{``mpg''} $\rightarrow$ \texttt{MPG}, \textit{``genre''} $\rightarrow$ \texttt{Major Genre}), (ii) semantic references, where words in a query are synonyms or semantically similar to the dataset attributes (e.g., \textit{``heavy''} $\rightarrow$ \texttt{Weight}, \textit{``fuel economy''} $\rightarrow$ \texttt{MPG}, \textit{``over time''} $\rightarrow$ \texttt{Year}), (iii) value-based references, where words in the query refer to cell values instead of column names (e.g., \textit{``1995 to 2010''} $\rightarrow$ \texttt{Release Year}, \textit{``furniture, office supplies, and technology''} $\rightarrow$ \texttt{Category}), and (iv) implicit references, where attributes are requested indirectly through a visualization type (e.g., if there is only one temporal attribute, requesting a line chart implies a reference to that attribute).
Chart type references include explicit requests for specific visualization types (e.g., \textit{``scatterplot''}, \textit{``bar chart''}, \textit{``histogram''}).
A subset of utterances also included explicit, unambiguous references to graphical encoding channels that should be bound to different attributes (e.g., \textit{``color by category,'' ``facet by origin''}).
Utterances with aggregation references included one or more words that mapped to the type of mathematical transform that needed to be applied to create a chart (e.g., sum, average, count).
Aggregation references were further sub-divided as explicit (e.g., \textit{``total gross''}->\texttt{SUM(Worldwide Gross)}, \textit{``number of orders''} $\rightarrow$ \texttt{COUNT}) or implicit via a requested chart type or phrasing (e.g., \textit{``histogram''} or \textit{``How many''} $\rightarrow$ \texttt{COUNT}, \textit{``stacked bar chart''} $\rightarrow$ \texttt{SUM} or \texttt{COUNT}).
Finally, utterances with explicit design references specified additional details such as mark color, chart orientation, sorting order, or axis tick windows that could be leveraged by visualization NLIs when rendering charts (e.g., \textit{``set origin colors to: europe, blue, japan, orange, usa, red,'' ``Histogram of running time, in 20 minute bands''}).

Note that the aforementioned phrasing and information types are by no means an exhaustive list, nor are they mutually exclusive.
For instance, while we tagged them as commands, one could cross-list the utterances \textit{``Faceting on Origin, plot Weight by Acceleration''} and \textit{``Can you create a graph showing sales and profit by region?''} as command+query and command+question, respectively.
Similarly, many utterances included keywords like \textit{`correlation,' `distribution,' `compare'} that could be used to identify intended analytic tasks~\cite{amar2005low}, which is another potentially relevant factor for a system when determining which visualizations to render.
Instead, the categorization described in this section is primarily to provide a subjective overview of the curated dataset and to serve as a reference point for more detailed discussions regarding the nature of utterances within visualization NLIs going forward.

%% file: sections/05-applications.tex
\section{Applications}
\label{sec:applications}

\added{The central contribution of this work is the curated dataset of 89\added{3} utterances (814 utterance sets) that are publicly available at: \projectURL.}
To lay the groundwork and help foster ideas for future research and development, below we highlight some exemplary applications of this dataset.


\subsection{Benchmarking Existing NL-based Visualization Tools}

\input{tables/nl4dv-benchmark}
The curated utterances can serve as a benchmarking dataset to evaluate the performance of existing systems.
As an example, we used the dataset to evaluate the NL4DV toolkit~\cite{narechania2020nl4dv} \added{which returns a list of Vega-Lite specifications in response to NL utterances}.
To benchmark NL4DV's performance, we configured NL4DV with the three datasets used for the study and executed the 755 singleton utterance sets curated from the study.
We only used singletons since NL4DV currently does not support sequential utterances.

Table~\ref{tab:nl4dv-performance} summarizes the toolkit's performance on the curated utterance dataset.
Execution times for individual utterances ranged between 0.01s--14.11s (mean: 1.04s)\footnote{Reported based on a MacBook Pro with a Dual-Core 2.6GHz processor and 8GB RAM running MacOS Catalina version 10.15.6}.
To compute the accuracy, we compared the list of Vega-Lite
specifications returned by NL4DV to the Vega-Lite chart shown during the study.
In our first iteration, we checked for exact matches between the mark type (\texttt{point}, \texttt{bar}, \texttt{line}), encodings (\texttt{x}, \texttt{y}, \texttt{color}, \texttt{column}), data bindings (i.e., mappings between attributes and encodings), and data aggregations (\texttt{count}, \texttt{mean}, \texttt{sum}) with one exception: we marked two scatterplots as ``equal'' even if their \texttt{x}, \texttt{y} data bindings were interchanged.
Inspecting the failed cases, we noticed that some mismatches were not interpretation or visualization recommendation errors but rather resulted from default data binding and aggregation choices made within NL4DV's recommendation engine (e.g., reversed \texttt{x} and \texttt{color} bindings for a stacked bar chart, setting a \texttt{mean} aggregation instead of a \texttt{sum} when no aggregation is specified in an utterance).
\added{Thus, to account for these default choices, we also computed a second ``partial match'' accuracy score ignoring such mismatches.}

Looking through the remaining failed utterances, we noticed that errors occurred due to NL4DV's keyword-based task detection logic (e.g., the word `relationship' would map to a \textit{correlation}, resulting in a \texttt{point} mark while the utterance was in the context of a bar chart), and undetected semantic attribute references (e.g., the phrase `\textit{fuel economy}' did not get mapped to the \textit{MPG} attribute), among others.
While not a critical review of NL4DV's performance, this example shows how the presented utterance dataset can help benchmark and identify areas for improvement within current tools.





\subsection{Developing New Models for NL-driven Data Visualization}

\input{tables/text-classification}
The curated utterance dataset can also be used to develop new techniques
to generate visualizations from NL.
As a preliminary application, we explored how the curated utterance dataset can be used to develop a classification model that predicts chart types (with labels being the ten chart types used in the study) based on NL utterances.
To this aim, we experimented with different classification techniques including logistic regression, random forests, support vector machines, and Naive Bayes available within Python's scikit-learn package~\cite{pedregosa2011scikit}.

We summarize the classification accuracy of these models trained on TF-IDF vectorized representations of the 814 utterance sets (singletons + sequential) in Table~\ref{tab:classification}.
Although we need substantially larger datasets to make these models practically useful, this initial setup illustrates the potential of the presented dataset for supporting future development.
Ultimately, such classifiers can be coupled with other NL understanding and visualization recommendation modules to generate visualizations in response to given utterances.
For example, one could combine the output from a classifier (i.e., chart type), detect attributes in input utterances with toolkits like NL4DV~\cite{narechania2020nl4dv}, and then determine the most perceptually effective bindings for the classified chart type's encoding channels using tools like CompassQL~\cite{wongsuphasawat2016towards} and Draco~\cite{moritz2018formalizing}.
Similarly, the curated utterances can also be used as input to contemporary few-shot learning models such as GPT-3~\cite{brown2020language}.
For instance, developers have already begun to experiment with GPT-3 for generating visualizations from NL using only tens of hand-crafted example utterances~\cite{twittergpt3}.
To this end, the presented dataset can help fuel these recent efforts and significantly improve performance by providing a broader training set covering multiple chart types and dataset domains.

%% file: tables/nl4dv-benchmark.tex
\begin{table}[t!]
\centering
\resizebox{\linewidth}{!}{%
\begin{tabular}{@{}lccc@{}}
\toprule
\multicolumn{1}{c}{\textbf{Utterances Used}} & \textbf{\begin{tabular}[c]{@{}c@{}}Accuracy\\ (Exact)\end{tabular}} & { \textbf{\begin{tabular}[c]{@{}c@{}}Accuracy\\ (Partial)\end{tabular}}} & \textbf{\begin{tabular}[c]{@{}c@{}}Execution\\ Time (Avg.)\end{tabular}} \\ \midrule
All (755) & 485 (64\%) & 569 (75\%) & 1.04 \\
Cars (290) & 203 (70\%) & 225 (78\%) & 0.21 \\
Movies (257) & 168 (65\%) & 204 (79\%) & 0.15 \\
Superstore (208) & 114 (55\%) & 140 (67\%) & 3.30 \\ \bottomrule
\end{tabular}%
}
\vspace{.5em}
\caption{NL4DV's accuracy and execution time (in seconds) for 755 singleton utterance sets from the curated dataset. \added{Note: results are based on a beta version of NL4DV 0.0.1}.}
\label{tab:nl4dv-performance}
\Description{NL4DV's performance summary on the curated dataset. Out of the 755 singleton utterance sets, 64\% of utterances yielded the exact chart shown during the study.}
\vspace{-2em}
\end{table}

%% file: tables/text-classification.tex
\begin{table}
\centering
\resizebox{.75\linewidth}{!}{%
\begin{tabular}{@{}lr@{}}
\toprule
{\textbf{Technique}} & {\textbf{Accuracy}} \\ \midrule
{Logistic Regression} & {88\%} \\
{Random Forest} & {88\%} \\
{Linear Support Vector Classification} & {87\%} \\
{Multinomial Naive Bayes} & {85\%} \\ \bottomrule
\end{tabular}%
}
\vspace{.5em}
\caption{Performance of different classification techniques using the 814 curated utterance sets. All numbers are reported based on a 10-fold cross validation.}
\label{tab:classification}
\Description{Accuracy of classification models trained on the curated dataset to predict intended chart types. All models had an accuracy of over 85\%, with Logistic Regression having the higest accuracy of 88\%.}
\end{table}

%% file: sections/06-discussion.tex
\section{Discussion}
\label{sec:discussion}

\subsection{Implications for System Design}

The spectrum of heavily specified utterances (e.g., \textit{``Show me a bar chart of the total profitability of each region with a breakdown based on shipping time''}) to highly underspecified keyword-based utterances (e.g., \textit{``Cylinders average mpg''}) highlights the varying level of expectations that people have from NLIs for data visualization.
Closely examining the collected utterances can help current systems expand their underlying grammar configurations to support more natural utterances, or help develop new systems based on the themes observed from this data.
As actionable takeaways, below we highlight some key design considerations for developing NLIs for visualization based on our observations from the curated set of utterances.

\subsubsection{Accommodating natural phrasings as part of user input in visualization tools.}
\begin{figure}
    \includegraphics[width=\linewidth]{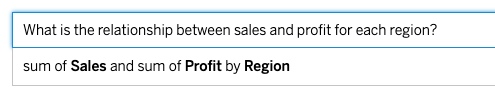}
    \caption{A phrasing suggestion in Tableau's Ask Data~\cite{tableauaskdata} while entering an utterance collected during the study.}
    \Description{Example illustrating how current NLIs for visualization enforce certain phrasing types through auto-complete features.}
    \label{fig:suggestions-askdata}
\end{figure}

While not an exhaustive or definitive list, the examples shown in Figure~\ref{fig:results}-top illustrate the diverse forms in which people naturally phrase NL utterances.
However, given practical implementation challenges, current NLIs often suggest utterances as users provide their input (e.g., Figure~\ref{fig:suggestions-askdata}), asking users to select from these utterances.
While these suggestions can aid discoverability and help improve system accuracy, their phrasing may mislead people about the system's interpretation capabilities or bias users towards phrasing their input in a specific format (e.g., query-like utterances as in the suggestion in Figure~\ref{fig:suggestions-askdata}).
Furthermore, mentally mapping different phrasings of input utterances and system suggestions can be tedious and potentially confusing (e.g., the input utterance in Figure~\ref{fig:suggestions-askdata} does not contain aggregations but the system suggestion does).
Hence, going forward, to enable truly ``natural'' language-based interactions and support flexible input phrasings, one consideration for visualization systems is to explore ways to internally translate input utterances into system-friendly phrasings without exposing these phrasings to users by default.

\subsubsection{Inferring different types of attribute references.}
Prior work by Grammel et al.~\cite{grammel2010information} has highlighted how people, especially novices, heavily rely on attributes to specify visualizations (deferring underlying visualization design decisions onto the system).
This observation has played a key role in motivating the design of NLIs for visualization as NL allows users (novices or experts) to freely specify attributes amidst naturally phrased questions or as part of targeted commands.
The collected utterances not only reaffirm this fact but also highlight the flexibility that NL affords in specifying attributes (explicit, semantic, value-based, and implicit references in Figure~\ref{fig:results}-bottom).
However, as also highlighted by prior work~\cite{gao2015datatone,setlur2019inferencing}, to be effective, NLIs must complement this input flexibility with robust interpretation techniques to infer these different references, particularly focusing on ambiguous (e.g., partial attribute matches) and semantic references (e.g., synonyms, domain-specific terminology such as \textit{``fuel economy"} for \texttt{MPG}).

\subsubsection{Balancing automated and manual view specification.}
Given the overarching task of visualization specification, an important aspect for supporting NL interaction is
making automated visualization design choices when an utterance is underspecified (e.g., choosing chart types based on attribute types, selecting a default aggregation when there is none specified in the query)~\cite{setlur2019inferencing}.
However, in this study, we observed that participants oftentimes provided explicit references to one or more encoding channels (e.g., \textit{``color by Origin,''} \textit{``faceted by region''}) that may not match a system's inherent design choices (e.g., a visualization system might select a colored scatterplot over a faceted one by default).
Yet, such custom specification is seldom supported in current systems that strive to provide agency through implicit decisions.
Thus, in line with recommendations from Tory and Setlur on prioritizing explicit intent during analytic conversations~\cite{tory2019mean}, in the context of visualization specification, future systems must accommodate explicit encoding requests, allowing users to override implicit visualization design choices made by current systems.

\subsection{Limitations and Future Work}


Although our work presents implications for the general design of NLIs for visualization, the collected data and our findings should be interpreted with the study's constraints and assumptions in mind.
Specifically, we only considered typed input and did not allow participants to use voice to enter utterances.
\added{Furthermore, we did not screen participants or collect any demographic information, and only distributed the study to participants outside the European region (\arjun{due to privacy laws and IRB requirements}).}
Hence, conducting follow-up studies to collect spoken utterances or targeting \added{focused user groups (e.g., visualization novices)} can help enrich the data and findings.

\added{As with most online studies, cleaning and filtering the data collected from the trials was an integral aspect of our work. For instance, even after incorporating the feedback from the pilot sessions, we still encountered a breadth of trials that had to be discarded (Section~\ref{sec:data-cleaning}). Furthermore, for individual trials, participants often entered two or more singleton utterance sets, one or more sequential utterance sets, or a combination of singleton and sequential utterance sets. While this highlights the input flexibility that the study interface provided, it also required us to manually inspect utterances and detect utterance sets. Given the relatively small scale of the data, performing these cleaning steps manually was feasible, albeit tedious. However, generalizing the data cleaning process for larger utterance datasets and investigating interactive data cleaning systems that function based on preconfigured rules or a set of examples are important areas for future work to explore.
}

Besides the immediate applications of the current dataset (Section~\ref{sec:applications}), our study can also be extended in different ways.
First, a short-term extension could be to consider \added{additional visualization types (e.g., parallel coordinates, heatmaps) or} focusing on non-tabular data (e.g., network visualizations, maps).
\added{Second, following the methodology used to construct popular NL-to-SQL datasets such as WikiSQL~\cite{zhong2017seq2sql}, a crowdsourced task could be setup to paraphrase the current set of utterances and then to validate the paraphrased utterances.
This paraphrasing can help generate a larger dataset which could subsequently enable more quantitative analyses of utterances and the development of NL-to-Visualization models, complementing research trends in the computer vision (e.g.,~\cite{antol2015vqa,das2017visual}) and database communities (e.g.,~\cite{yu2019spider,yu2019cosql}).}
Lastly, we asked participants to specify only one chart at a time.
However, NLIs have the potential to offer a more fluid analytic experience which goes beyond specifying a single chart.
To this end, future studies could combine ideas from the current study and prior research on conversational visual analysis~\cite{tory2019mean,hoque2017applying} to collect data on how people specify dashboards or sequences of charts (as opposed to specifying each chart from scratch).

%% file: sections/07-conclusion.tex
\section{Conclusion}


We conducted an online study with 102 participants to collect NL utterances people use to specify data visualizations.
To guide future research and development, we make the curated dataset of 89\added{3} utterances publicly available at \projectURL~and illustrate its application toward the creation and benchmarking of NLIs for data visualization.
In this paper, providing insight into the nature of these utterances, we characterize them based on their phrasing type (e.g., commands, queries, questions) as well as the visualization specification-relevant information they provide (e.g., chart types, encodings, aggregation functions).
\added{Additionally, we present the system design implications of the observed utterance patterns and briefly reflect on design considerations for conducting subsequent online studies in the space.
Finally, we discuss the constraints of our study along with future research opportunities that can complement our study, enriching the data and findings.}

%% file: sections/acknowledgement.tex
\begin{acks}
\added{We thank our study participants for their time and effort, and the anonymous reviewers for the detailed and helpful feedback on the article. This work was supported in part by a National Science Foundation Grant IIS-1717111.}
\end{acks}